\documentclass[twoside,11pt]{9ssmmp} %%% Please do not change this class !!! %%%
\usepackage{fancyhdr}
\usepackage{amssymb,amstext,amsmath}

\newcommand{\D}{\Delta}

\renewcommand{\L}{\Lambda}
\renewcommand{\l}{\lambda}
\renewcommand{\S}{\Sigma}
\newcommand{\G}{\varGamma}
\newcommand{\g}{\gamma}
\newcommand{\e}{\epsilon}
\newcommand{\s}{\sigma}
\renewcommand{\o}{\omega}
\renewcommand{\O}{\Omega}

\renewcommand{\a}{\alpha}
\renewcommand{\b}{\beta}

\newcommand{\m}{\mu}
\newcommand{\n}{\nu}
\renewcommand{\r}{\rho}
\renewcommand{\s}{\sigma}

\newcommand{\f}{\phi}

\renewcommand{\th}{\theta}

\renewcommand{\e}{\epsilon}

\renewcommand{\det}{\mathop{\rm det}\nolimits}

\newcommand{\be}{\begin{equation}}
\newcommand{\ee}{\end{equation}}
\newcommand{\bea}{\begin{eqnarray}}
\newcommand{\eea}{\end{eqnarray}}

\begin{document}

%% Please modify the following line to include the title of your contribution and acknowledgments:

\title{Quantum gravity for piecewise flat spacetimes\hspace{.25mm}\thanks{\,This work has been supported by GFMUL (Mathematical Physics Group at University of Lisbon) and the project ON 171031 of the Ministry of Education, Science and Technological Development, Serbia.}}

%% Please modify the following lines to include author names, affiliations and e-mail addresses:

\author{\bf{Aleksandar Mikovi\'c}\hspace{.25mm}\thanks{\,e-mail address: amikovic@ulusofona.pt} \\
\normalsize{Departamento de Matem\'atica,
Universidade Lus\'ofona} \\ \normalsize{Av. do Campo Grande, 376, 1749-024 Lisboa, Portugal} \vspace{2mm} \\
\bf{Marko Vojinovi\'c}\hspace{.25mm}\thanks{\,e-mail address: vmarko@ipb.ac.rs} \\
\normalsize{Institute of Physics, Pregrevica 118, 11080 Belgrade, Serbia}}

\date{} %% Please do not modify

\maketitle %% Please do not modify

\begin{abstract}
We describe a theory of quantum gravity which is based on the assumption that the spacetime structure at small distances is given by a piecewise linear (PL) 4-manifold corresponding to a triangulation of a smooth 4-manifold. The fundamental degrees of freedom are the edge lengths of the triangulation. One can work with finitely many edge lengths, so that the corresponding Regge path integral can be made finite by using an appropriate path-integral measure. The semi-classical limit is computed by using the effective action formalism, and the existence of a semi-classical effective action restricts the choice of the path-integral measure. The classical limit is given by the Regge action, so that one has a quantum gravity theory for a piecewise-flat general relativity. By using the effective action formalism we show that the observed value of the cosmological constant can be recovered from the effective cosmological constant. When the number of 4-simplices in the spacetime triangulation is large, then the PL effective action is well approximated by a quantum field theory effective action with a physical cutoff determined by the smallest edge length. 
\end{abstract}

\section{Introduction}

The standard approach to the problem of constructing a quantum gravity (QG) theory \cite{I,I2} can be described as the following problem. Let $M$ be a smooth 4-manifold, of topology $\Sigma\times I$, where $\Sigma$ is a 3-manifold and $I$ an interval from $\bf R$. Let $g$ be a Minkowski-signature metric on $M$  and $\Phi$ a set of matter fields on $M$. Then the goal is to find a triple $(\hat g, \hat\Phi, \hat U )$, where $\hat g$ and $\hat\Phi$ represent Hermitian operators parametrized by the points of $M$,  acting in some Hilbert space $\cal H$, while $\hat U$ is a unitary evolution operator parametrized by $I$, such that the  the classical limit ($\hbar\to 0$) of the quantum time-evolution is equivalent to the Einstein equations.

The best known example of this approach is Loop Quantum Gravity (LQG), see \cite{lqg} for a recent review and references. In the LQG case, the Hilbert space $\cal H$ is only known to be a subset of a non-separable Hilbert space and $\hat U$ can be constructed only for a triangulation $T(M)$ of $M$, so that it is not clear what is the classical limit. Note that in the standard QG approach, the structure of $M$ is not changed after the quantization, and it is well known that this is the main source of the difficulties for a quantization of gravity \cite{I,I2}. This leads us to an alternative approach where $M$ is replaced by a quantum spacetime $\widehat{M}$. The obvious choice would be a non-commutative manifold based on $M$, like in the case of noncommutative geometry (NCG) \cite{ncg}, where the coordinates of $M$ become elements of a noncommutative algebra. Another choice is made in the superstring theory \cite{sst}, where the coordinates of $M$ become coordinates of the loop manifold ${\cal L}M$ and new Grassmann (anticommuting) coordinates are added, so that $\widehat M$ is a loop super manifold.

In this paper we would like to present the case when $\widehat{M} = T(M)$, see \cite{mik, mv}. This is clearly a much simpler choice for $\widehat M$ than the one made in NCG or in the superstring theory, but the price paid is that the spacetime triangulation becomes a physical structure. However, the PL manifold $T(M)$ looks like the smooth manifold $M$ when the number of 4-simplices is large. Also, by using $T(M)$ one reduces the infinite number of the degrees of freedom (DOF) for $g$ and $\Phi$ to a finite number, which then simplifies the quantization. 

Note that Regge was the first to use $T(M)$ in order to define the path integral for general relativity (GR) \cite{r}, see \cite{h} for a modern review. However, in Regge's approach the triangulation was an auxiliary structure and had to be removed via the smooth limit $T(M) \to M$. However, obtaining the smooth limit in the Regge approach is a difficult problem. The same applies to the case of spin-foam models of LQG, which can be only defined when the spacetime is a PL manifold. In causal dynamical triangulations (CDT) approach \cite{agjl}, $T(M)$ is also used to define the path integral, but it is also considered an auxiliary structure. Obtaining the smooth limit in CDT is proposed by performing a sum over the triangulations.

\section{PL gravity path integral}

Let $T(M)$ be a regular\footnote{Any two $k$-simplices of $T(M)$ cannot have more than one common $(k-1)$-simplex, where $k=1,2,3,4$.} triangulation of a smooth 4-manifold $M$. We will assign positive numbers $L_\epsilon$ to the edges $\epsilon$ of $T(M)$. If we think of an $L_\e$ as a distance between two vertices of $T(M)$ induced by some metric, then we can define a constant metric in each 4-simplex $\sigma$
\begin{equation}
g_{kl}^{(\sigma)} = \frac{ L_{0k}^2 + L_{0l}^2 -L_{kl}^2}{2L_{0k} L_{0l}} \,, \qquad 1\le k,l \le 4 \,,\label{plm}
\end{equation}
where the indices $0,1,2,3,4$ denote the vertices of a 4-simplex $\s$. Hence we replace a smooth metric $g$ on $M$ by a PL version (\ref{plm}). We want that the PL metric has the Minkowski signature, and this can be ensured by requiring that $L_\e$ satisfy the triangle inequalities for the triangles which belong to one of the tetrahedrons of $\s$, for example the tetrahedron $(1,2,3,4)$, while the $L_\e$ of the triangles $(0,i,j)$ must not satisfy the triangle inequalities. 

Having all $L_\e > 0$ means that all triangles in $T(M)$ are spacelike. For $M = \S \times I$ manifolds, this gives an accordion-like triangulation (triangulation of a cylinder). A more natural triangulation is to take a finite number of spacelike slices $T_k (\S)$ which are linked by timelike edges such that each 4-simplex has a spacelike tetrahedron in $T_k$ and a vertex in $T_{k-1}$ or in $T_{k+1}$. This class of triangulations is used in CDT models \cite{agjl}. We will then require that the $L_\e$ of $T_k$ satisfy the triangle inequalities, while a timelike edge will be assigned an imaginary length $i L_\e$. Hence the labels of the edges of timelike triangles will not satisfy the triangle inequalities and the metric (\ref{plm}) will have the correct signature.

The curvature scalar $R$ will be concentrated on the triangles and $R$ will be given by the deficit angle divided by the area of the dual face. Hence in each $\sigma$ we have a flat metric (\ref{plm}) so that we can say the corresponding PL metric is a piecewise-flat metric. 

Note that an $L_\e$ label represents a proper length, so that $L_\e$ is invariant under the local Lorentz transformations in each 4-simplex. We will also have $(L_\e)^2 > 0$ for a spacelike edge, while $(L_\e)^2 < 0$ for a timelike edge. 

The Einstein-Hilbert action for the PL metric (\ref{plm}) becomes the Regge action
\be S_{Rc} = \frac{1}{G_N}\sum_{\D=1}^F  A_\D (L) \theta_\D (L) + \L_c  V_4 (L) \,,\label{rac}\ee 
where $G_N$ is the Newton constant, $A_\D (L)$ is the area of a triangle $\D\in T(M)$ and $\th_\D$ is the deficit angle. $\L_c$ is the cosmological constant and $V_4$ is the 4-volume of $T(M)$. See \cite{agjl} how to define (\ref{rac}) when the timelike triangles are present. Note that the Regge action describes a theory with a finite number of DOF when $\S$ is compact, while in the case when $\S$ is non-compact, we can restrict $L_\e$ to be non-zero only in a ball $B\subset\Sigma$.

One can also couple the matter fields to a Regge PL metric and the corresponding smooth actions will become the PL actions for a finite number of matter DOF. For example, a scalar field will be defined by the values of the field at the vertices of $T(M)$, which is equivalent to a PL function on the 4-polytopes of the dual triangulation.

In the case of a scalar field matter, the Regge path integral will be given by the following $(E+V)$-dimensional integral 
\be Z = \int_{ D_{E}} \, \mu (L) \, d^E L \,\int_{{\bf R}^V} \prod_{\nu=1}^V d\f_\nu \, e^{ i [S_{Rc}(L) +  S_m (L,\f)]/\hbar} \,, \label{crss}\ee
where $E$ is the number of the edges in $T(M)$ and $V$ is the number of the vertices in $T(M)$ \cite{mv}. $S_m$ is the PL form of the scalar-field action and the integration region $D_E$ is a subset of ${\bf R}_+^E$ where the triangle inequalities hold. The measure $\m$ has to be chosen such that it makes $Z$ finite. The matter PI measure is taken to be trivial and we will assume that the matter path integral is finite. This is true, because the matter path integral will be given by a finite product of the  integrals of the type
\be  I(\a,\b) = \int_{-\infty}^\infty dx\, e^{-\a x^2 - \b x^4} \,,\ee
where $\a,\b \in \bf C$. Since $I$ is convergent for $\a,\b > 0$, the analytic continuation $I(i\a,i\b)$ will be finite. 

Note that in the standard Regge formulation the spacetime metric is of the Euclidean signature. This was done in analogy to the QFT case where the Euclidean signature improves the convergence of the path integral (\ref{crss}). However, in the QG case this does not help, because the scalar curvature also changes the sign in the Euclidean case and can be unbounded. Actually, the Lorentzian integral has better convergence properties, which can be seen on a toy example $R(x) = \a\, x^2$ where $x\in\bf R_+$ and $\a$ is a constant different from zero. Then $Z_E = \int_0^{\infty} dx\, e^{- R(x)}$ is convergent only for $\a > 0$ while $Z_L = \int_0^\infty dx\, e^{iR(x)}$ is convergent for any sign of $\a$. The presence of imaginary edge lengths and imaginary angles in the Lorentzian case is not a problem, since all the geometric quantities can be defined \cite{agjl}.

Finding the smooth limit $T(M)\to M$ for $Z$ is a difficult problem. However, there is a promising approach, based on the Wilson renormalization group \cite{h}. In this approach one considers $Z$ as function of the dimensionless couplings $\g$ e $\l$
$$\g  =  l_0^2/(G_N \hbar) =  l_0^2 /l_P^2 \,,\quad \l = l_0^4\L_c / \hbar = l_0^4/ (L_c^2 l_P^2)\,, $$
where $L_c^2 = G_N /\L_c$ and $l_0$ is an arbitrary length. One then looks for a critical point $P_0 = (\g_0 ,\l_0)$ where the second derivatives of $Z$ diverge so that there is a second-order phase transition. At the critical point the correlation length diverges, so that a transition to the smooth phase occurs. However, the problem with this approach is that at $P_0$ the perturbation theory does not apply, so that the calculation has to be done by using numerical methods. Also the semiclassical limit $l_P^2 \to 0$ corresponds to a strong coupling region $\g\to \infty$ and $\l \to \infty$ so that it is difficult to determine it analytically.

However, the easiest way to determine the semiclassical limit in a QG theory defined by a path integral is to use the effective action, see \cite{mv1,mv2, msc,mik,mv}. Namely, the effective action can be calculated analytically in the $\hbar\to 0$ limit. Also the PI measure $\m(L)$ has to be such that allows a semiclassical expansion for the effective action for large $L_\e$. This gives us an additional constraint on the choice of $\m(L)$.

\section{Effective action for PL quantum gravity}

We will assume that $T(M)$ is the fundamental spacetime structure, i.e. the spacetime is a  {\it piecewise linear} 4-manifold $T(M)$ with a flat metric in each cell (4-simplex $\s$). If $N$ is the number of cells of $T(M)$, then for $N \gg 1$, $T(M)$ will look like the smooth manifold $M$ on a scale much larger than the maximal edge length. 

By an appropriate choice of the measure $\m$ the integral $Z(T(M))$ can be made finite. Since $T(M)$ is the physical spacetime, there is no need to define the smooth limit $T(M)\to M$. Instead, we need a large-$N$ approximation for the observables. This is analogous to the fluid dynamics situation where on the scales much larger than the inter-molecular distance we can approximate the molecular velocities as a smooth field and use the Navier-Stokes equations.

We will determine the semiclassical limit of PL quantum gravity by using the effective action. It can be computed by using the effective action equation in the limit $L_\epsilon \gg l_P = \sqrt{G_N  \hbar}$.

Let us recall first the effective action definition from quantum field theory (QFT). Let $\phi$ be a real scalar field on $M$ and let
$$ S (\phi) = \frac{1}{2}\int_M d^4 x \sqrt{|g|}\left[g^{\mu\nu}\,\partial_\mu \phi \,\partial_\nu \phi -  \frac{1}{2}\o^2  \phi^2 - \lambda\,\phi^4 \right]\,,$$
be a flat-spacetime action. The effective action $\varGamma (\phi)$  can be determined from the following integro-differential equation
\be e^{i\varGamma(\phi)/\hbar}=\int {\cal D} h \exp\left[\frac{i}{\hbar} S(\phi + h) -\frac{i}{\hbar} \int_M d^4 x\,\frac{\delta\varGamma}{\delta\phi (x)}h(x)\right]\,, \ee
see \cite{n,k}.

Note that a generic solution $\varGamma(\phi)$ is a function with values in ${\bf C}$. The Wick rotation is used to obtain a real-valued function $\varGamma(\phi)$. This is done by solving first the EA equation in the Euclidean spacetime
\be e^{-\varGamma_E (\phi)/\hbar}=\int {\cal D} h \exp\left[-\frac{1}{\hbar} S_E (\phi + h) +\frac{1}{\hbar} \int_M d^4 x\,\frac{\delta\varGamma_E}{\delta\phi (x)}h(x)\right]\,.\ee
Then $x_0 = - it$ is inserted into a solution $\varGamma_E (\phi)$, where $(x_0, x_k)$ are the spacetime coordinates, so that 
$$\G (\phi) = i\G_E (\phi)|_{x_0 = -it} \,.$$

However, the Wick rotation cannot be used in quantum gravity, since in many problems of interest, introducing a flat background metric does not make sense. One way to resolve this difficulty is to use the fact that the Wick rotation in QFT is equivalent to 
\be \varGamma(\phi) \to Re\,\varGamma(\phi) + Im\,\varGamma(\phi)\,,\ee
see \cite{mv1,mv2}. This prescription is convenient for quantum gravity because it does not involve a background metric, nor a system of coordinates.

In the case of PL quantum gravity without matter, the effective action (EA) equation is given by 
\be e^{i{\varGamma}(L)/l_P^2} = \int_{D_E(L)} d^E  x \, \mu (L +x) e^{iS_{Rc} (L+x)/l_P^2 - i\sum_{\epsilon=1}^E {\varGamma}'_\epsilon (L)x_\epsilon /l_P^2 } \,,\ee
where $l_P^2 = G_N\hbar$ and $D_E(L)$ is a subset of ${\bf R}^E$ obtained by translating $D_E$ by a vector $-L$ \cite{mik}. Note that $D_E (L) \subseteq  [-L_1, \infty)\times\cdots\times[-L_E, \infty)$.

We will look for a semiclassical solution
$$ \varGamma (L) = S_{Rc} (L) + l_P^2 \varGamma_1 (L) + l_P^4 \varGamma_2 (L) + \cdots \,,$$
where $L_\epsilon \gg l_P$ and
$$| \varGamma_n (L) | \gg l_P^2 |\varGamma_{n+1}(L) |\,. $$

When $L_\epsilon \to \infty$, then $D_E(L) \to {\bf R}^E$ and
\be e^{i{\varGamma}(L)/l_P^2} \approx \int_{{\bf R}^E} d^E x \, \mu (L +x) e^{iS_{Rc} (L+x)/l_P^2 - i\sum_{\epsilon=1}^E {\varGamma}'_\epsilon (L)x_\epsilon /l_P^2 } \,.\label{llea}\ee

Actually, one can use the equation (\ref{llea}) to determine $\G(L)$ for large $L$ when $\m$ falls off sufficiently quickly \cite{mik}. The reason is that 
$$D_E (L) \approx [-L_1,\infty)\times\cdots\times[-L_E,\infty)\,,$$
for $L_\e \to \infty$, so that the relevant behaviour is captured by the following one-dimensional integral
$$
\int_{-L}^\infty dx \, e^{-zx^2/l_P^2 - wx} = \sqrt{\pi}\,l_P \exp{\Big [}-\frac{1}{2} \log z + l_P^2 \frac{w^2}{4z}
$$
$$
+l_P\frac{e^{-z\bar L^2 /l_P^2}}{2\sqrt{\pi z}\bar L}\left(1 + O(l_P^2 /z\bar L^2)\right) {\Big ]} \,,
$$
where $\bar L = L + l_P^2 \frac{w}{2z}$ and $Re\, z =- (\log\mu )'' $. The non-analytic terms in $\hbar$ will be absent if
 $$ \lim_{L\to\infty}e^{-z\bar L^2 /l_P^2} = 0 \Leftrightarrow (\log\m)'' < 0 \,\,\textrm{for}\,\, L\to\infty\,. $$
Hence the perturbative solution exists for the exponentially damped measures and it will be given by the equation (\ref{llea}).

For $D_E(L) ={\bf  R}^E$ and $\mu(L)$ a constant, the perturbative solution is given by the EA diagrams
$$ \varGamma_1 =\frac{i}{2}Tr \log S''_{Rc}  \,,\quad \varGamma_2 = \langle S_3^2 G^3 \rangle + \langle S_4 G^2 \rangle\,, $$
and
$$ \varGamma_3 = \langle S_3^4 G^6 \rangle + \langle S_3^2 S_4 G^5  \rangle + \langle S_3 S_5 G^4  \rangle + \langle S_4^2 G^4  \rangle + \langle S_6 G^3 \rangle \,, \,... $$
where $G = i(S_{Rc}'')^{-1}$ is the propagator and $S_n = iS_{Rc}^{(n)}/n!$ for $n > 2$, are the vertex weights, see \cite{k,mik}. The contractions $\langle X \cdots Y \rangle$ are the sums over the repeated DOF indices
$$\langle X\cdots Y\rangle =  \sum_{k,...,l} X_{k...l} \cdots Y_{k...l}\quad. $$

When $\mu(L)$ is not a constant, then the perturbative solution is given by
$$ \varGamma (L) = \bar S_{Rc} (L) + l_P^2 \bar\varGamma_1 (L) + l_P^4 \bar \varGamma_2 (L) + \cdots \,,$$ 
where 
$$\bar S_{Rc} = S_{Rc} - il_P^2 \log\mu \,,$$
while $\bar\varGamma_n$ is given by the sum of $n$-loop EA diagrams with $\bar G$ propagators and $\bar S_n$ vertex weights \cite{mik}.

Therefore
$$
\varGamma_1 = -i\log\mu + \frac{i}{2}Tr \log S''_{Rc} \,,
$$
$$
\varGamma_2 = \langle S_3^2 G^3 \rangle + \langle S_4 G^2 \rangle + Res [l_P^{-4} Tr\log\bar G ]\,,
$$
$$
\varGamma_3 = \langle S_3^4 G^6 \rangle + \cdots + \langle S_6 G^3 \rangle + Res [l_P^{-6} Tr\log\bar G ] 
$$
$$
+ Res[l_P^{-6}\langle \bar S_3^2 \bar G^3 \rangle] + Res[l_P^{-6}\langle \bar S_4 \bar G^2 \rangle]\,,
$$
see \cite{mik}.

Since the PI measure $\m$ has to vanish exponentially for large edge lengths, a natural choice is
\be \mu(L) = \exp\left( - V_4 (L) /(L_0)^{4} \right)\,, \label{ccm}\ee
where $L_0$ is a length parameter \cite{mik}. Since $\log\mu(L) = O\left( (L /L_0)^{4} \right)$\footnote{The notation $f(x_1,...,x_n) = O(x^\a)$ means that $f(\l x_1,...,\l x_n) = O(\l^\a)$ for $\l\to\infty$.} then for 
$L_\epsilon > L_c$ and
\be L_0 > \sqrt{l_P\, L_c}\,,\label{pecc}\ee  
where $L_c^{-2} = \L_c$, we get the following large-$L$ asymptotics \cite{mv,mvp}
\be\varGamma_1 (L) = O(L^4/L_0^4) + \log O(L^2/L_c^2) + \log\theta(L) + O(L_c^2 /L^2) \label{go}\ee
and
\be  \varGamma_{n+1}(L) = O\left((L_c^2 /L^4)^{n}\right) + L_{0c}^{-2n} O\left((L_c^2 /L^2)\right)\,, \label{gn}\ee
where $L_{0c} = L_0^2 /L_c $.

\section{Effective cosmological constant}

The asymptotics (\ref{go}) and (\ref{gn})  imply that the series 
$$\G(L) = \sum_{n\ge 0} ( l_P)^{2n}\varGamma_n (L)$$
is semiclassical (SC) for $L_\epsilon \gg l_P$ and  $ L_0 \gg \sqrt{l_P\, L_c}$.

Let $\varGamma \to \varGamma /G_N$ so that $ S_{eff} = ( Re\,\varGamma + Im\,\varGamma) /G_N$.
The effective action is then given by
$$
S_{eff} = \frac{S_{Rc}}{G_N} + \frac{l_P^2}{G_N L_0^4} V_4 + \frac{l_P^2}{2G_N} Tr\log S''_{Rc}  + O(l_P^4)\,,
$$
for $L_\e \gg l_P$. Hence  the $O(\hbar)$, or the one-loop, cosmological constant (CC) for pure gravity is given by
\be \Lambda = \L_c + \frac{l_P^2}{L_0^4} = \L_c + \L_{qg} \,.\label{qgcc}\ee

One can show that the one-loop cosmological constant is exact because there are no $O(L^4)$ terms beyond the one-loop order \cite{mv,mvp}. This is a consequence of the large-$L$ asymptotics
$$ \log \bar S''_{Rc}(L) = \log O(L^2 /\bar L_c^2) + \log\theta(L) +O(\bar L_c^2 /L^2) $$
$$\bar\G_{n+1} (L) = O\left((\bar L_c^2 /L^4)^n \right) \,,$$
where $\bar L_c^2 = L_c^2 \left[ 1 + il_P^2 ( L_c^2 /L_0^4 )\right]^{-1/2}$.

Hence the one-loop formula (\ref{qgcc}) is exact in the case of pure gravity.
If $\L_c =0$, the observed value of $\Lambda$ is obtained for $L_0 \approx 10^{-5} m$ so that $ l_P^2 \Lambda \approx 10^{-122}$ \cite{mik}. Note that $L_0 \approx 10^{-5} m$ is consistent with the requirement that $L_0 \gg l_P$, which replaces the SC condition $L_0 \gg \sqrt{L_c l_P}$ when $\L_c = 0$. 

The formula (\ref{qgcc}) is intriguing but unrealistic, since there is matter in the universe. In order to obtain a realistic expression for the effective CC, we need to study the EA equation with matter. This study also requires the understanding of the emergence of the smooth spacetime from a PL manifold $T(M)$. If $T(M)$ has a large number of the edges ($E \gg 1$) then the following approximations are valid
\be S_{R} (L) \approx \frac{1}{2} \int_M d^4 x \sqrt{|g|} \, R(g) \,,\ee
and 
\be \Lambda_c V_4 (L) \approx \Lambda_c \int_M d^4 x \sqrt{|g|} = \L_c \,V_M \,, \ee
where $|g| = |\det \,g|$. These are the standard formulas of the Regge calculus and they nicely illustrate how the PL manifold $T(M)$ with many 4-simplices can be approximated by a smooth manifold $M$ with a smooth (differentiable) metric $g$.

Similarly, the effective action $\G(L)$ will be approximated by a QFT effective action $\G^* (g)$, where $g$ is a smooth metric on $M$. Let $L_K$ be a minimal length in a triangulation, so that $L_\epsilon \ge L_K$ and let $L_K \gg l_P$. When $E\gg 1$ the following approximation is valid
\be Tr \log S_R''(L) \approx \int_M d^4 x \sqrt{|g|} \left[ a R^2 + b R_{\mu\nu}R^{\mu\nu}\right]\log(K/K_0)\,,\label{ola}\ee
where $R_{\mu\nu}$ is the Ricci tensor, and $a,b,K_0$ are some constants.

The formula (\ref{ola}) follows from the fact that a PL function on a lattice with a cell size $L_K$ can be written as a Fourier integral over a compact region $|q|\le \pi/L_K$ where $q$ is the wave vector\footnote{This region is known as the first Brillouin zone.}. Hence the PL trace-log term can be approximated by using the QFT formulation of GR with a momentum cutoff $K = 2\pi\hbar/L_K$.
 
The effect of the matter on the CC can be studied by introducing a scalar field on $M$
\be S_m (g,\phi) = \frac{1}{2}\int_M d^4 x \sqrt{|g|}\left[g^{\mu\nu}\,\partial_\mu \phi \,\partial_\nu \phi -  U(\phi) \right]\,,\label{scfa}\ee
where $U=\frac{1}{2}\o^2  \phi^2 + \lambda\,\phi^4$.
 
On a PL manifold $T(M)$ the action (\ref{scfa}) becomes
$$ S_{m} = \frac{1}{2}\sum_\sigma V_\sigma (L)  \sum_{k,l} g^{kl}_\sigma (L)\,  \phi'_k \, \phi'_l - \frac{1}{2}\sum_p V_p^* (L)\, U( \phi_p) \,,$$
where $\phi'_k = (\phi_k -\phi_0 ) /L_{0k}$ and $k,l,0$ are vertices in a 4-simplex $\s$, $p$ labels the vertices of $T(M)$ and $V^*$ is the volume of the dual cell. Then the total classical action of gravity plus matter on $T(M)$ is given by
$$ S(L,\phi) =\frac{1}{G_N} S_{Rc}(L) + S_m (L,\phi) \,. $$

The corresponding EA equation is given by
$$
e^{\frac{i}{l_P^2}\varGamma (L,\phi)} = \int_{D_E (L)} d^E l \,\int_{{\bf R}^V}\,d^V \chi \hphantom{mmmmmmmmmmmmmmmm}
$$
\be
\hphantom{mmmmm}
 \exp \left[ \frac{i}{l_P^2} \left( \bar S (L+l, \phi + \chi)
-\sum_\epsilon \frac{\partial\varGamma}{\partial L_\epsilon }\,l_\epsilon -\sum_p \frac{\partial\varGamma}{\partial \phi_p }\,\chi_p \right) \right] \,,\label{gmea}
\ee
where $\bar S = S_{Rc} - i l_P^2 \log \m  + G_N S_{m}$, see \cite{mv}.

We will look for a perturbative solution
$$ \G(L,\f) = S(L,\f) + l_P^2 \G_1 (L,\f) + l_P^4 \G_2 (L,\f) + \cdots \,,$$
and require it to be semiclassical for $L_\e \gg l_P$  and $|\sqrt{G_N}\,\f | \ll 1$. This can be checked on the $E=1$ toy model
$$S(L,\f) =( L^2 + L^4/L_c^2 )\theta(L) + L^2\theta(L)\f^2 (1+  \o^2 L^2 + \lambda\f^2 L^2)\,,$$
where $\theta(L)$ is a homogeneous function of degree zero.

It is not difficult to see that 
$$\G (L,\f) = \G_g (L) + \G_m (L,\f)\,,$$
and 
$$\G_m (L,\f) = V_4 (L) \, U_{eff} (\f)$$
for constant $\f $ where $U_{eff}(0) =0$. Furthermore,
$$\G_g (L) = \G_{pg}(L) + \G_{mg}(L) \,,$$ 
where $\G_{pg}$ is the pure gravity contribution and $\G_{mg}$ is the matter induced contribution. 

In the smooth-manifold approximation one has
$$ \G_{mg}(L) \approx \L_m V_M + \Omega_m (R,K) \,, $$
where $K=2\pi\hbar/L_K$ is the momentum cutoff. One can show that 
$$\O_m = \O_1 l_P^2 + O( l_P^4)$$
and
\be
\label{krun}
\begin{array}{lcl} 
\O_1 (R,K) &=& \displaystyle a_1 K^2 \int_M d^4 x \sqrt{|g|}\,R \\
 &+& \displaystyle \log(K/\o)\, \int_M d^4 x \sqrt{|g|}{\Big [ }a_2 R^2 + a_3 R^{\m\n}R_{\m\n} \\
& & \displaystyle \hphantom{mmmmmmmmmmm} + a_4 R^{\m\n\r\s} R_{\m\n\r\s}+ a_5\nabla^2 R {\Big ]}\\
&+& \displaystyle O(1/K^2) \,, \\
\end{array}
\ee
where $R_{\m\n\r\s}$ is the Riemann curvature tensor, see \cite{mv}.

The effective CC will be then given as
$$\L =\L_c +\L_{qg} +\L_m \,,$$
where $\L_{qg}$ is given by (\ref{qgcc}). Note that the matter contribution to CC can be approximated by a sum
\be \L_m \approx \sum_\g v(\g, K) \ee 
where $v(\g,K)$ is a one-particle irreducible vacuum Feynman diagram for the field-theory action $S_m$ in flat spacetime with the  cutoff $K$. One can show that 
\bea \sum_\g v(\g, K) &\approx & l_P^2 \, K^4 {\Big [} c_1 \ln (K^2/\o^2)+ \sum_{n\ge 2}c_n  (\bar\l)^{n-1}(\ln (K^2/\o^2))^{n-2} \cr
 &+& \sum_{n\ge 4} d_n(\bar\l)^{n-1}( K^2/\o^2 )^{n-3} {\Big ]}\,,\label{qftcc}\eea
for $K \gg \omega$, where $\bar\l = l_P^2 \l$, see \cite{mvp}. Therefore one has a highly divergent sum of matter vacuum-energy contributions to the cosmological constant when $K\to\infty$. This is the famous cosmological constant problem which appears in any QFT formulation of quantum gravity.

However, in the PL formulation of quantum gravity (PLQG), the QFT which produces the infinite sum in (\ref{qftcc}) is just an approximation. The fundamental theory has finitely many DOF so that the exact solution of the EA equation will give a finite and cutoff-independent value for $\L$. Therefore  
\be \L_m = V( \o^2 , \lambda,l_P^2) \,,\ee 
and
\be\L = \pm \frac{1}{L_c^2} + \frac{l_P^2}{2L_0^4} + V(\o2 , \lambda, l_P^2 )\,. \label{efcc}\ee

The equation (\ref{efcc}) can be used to fix the free parameters $L_0$ and $L_c$. By equating $\L$ with the experimentally observed value, we obtain
\be \l = x + y + \l_m \label{pe}\ee
where $\l = l_P^2 \L \approx 10^{-122}$, $x = \pm\, l_P^2 / L_c^2$, $y =l_P^4 / 2L_0^4$ and $\l_m = l_P^2 V$. The equation (\ref{pe}) has infinitely many solutions, but we also have to impose the condition for the existence of the semi-classical limit (\ref{pecc}). This gives the restriction 
\be 0 < y < 2|x| \,.\label{scr}\ee 

The value of $\l_m$ is not known, but for any value of $\l_m$ the equation (\ref{pe}) has infinitely many solutions which obey the restriction (\ref{scr}). Note that the solution $x = -\l_m$ and $y = \l$, which was proposed in \cite{mv}, will be acceptable if $|\l_m | > \l / 2$. This solution is special because it gives a value for $L_0$ which is independent of the value of $\l_m$, $L_0 \approx 10^{-5}$m. This is the same value which was obtained in the case of pure PL gravity without the cosmological constant \cite{mik}.

\section{The CC problem in quantum gravity}

The formula (\ref{efcc}) for the exact effective cosmological constant is an essential ingredient for the resolution of the CC problem from QFT in the context of a QG theory. The result (\ref{efcc}) can be better understood if we recall the definition of the CC problem given by Polchinski \cite{pol}. According to this definition, the CC problem in a QG theory has two parts:

1) show that the observed CC value is in the CC spectrum,

2) explain why the CC takes the observed value.

The meaning of the first part (P1) of the CC problem is obvious if the cosmological constant is represented by an operator. In the case when one has a quantum corrected expression of the classical CC value, one has to show that there are values of the free parameters which give the observed CC value. The PLQG theory clearly solves P1, while the second part (P2) of the CC problem cannot be addressed by the current formalism. The reason is that one has to generalise the standard formalism of quantum mechanics in order to provide a mechanism for a selection of a wavefunction of the universe with a particular value of the cosmological constant. 

Note that demonstrating P1 is a highly non-trivial task in any QG theory. The problem P1 has been addressed so far only in PLQG theory and in string theory. In the string theory case there are only plausibility arguments that P1 is true \cite{rp,kklt}. The CC spectrum in string theory is discrete with $O(10^{500})$ values \cite{rp}. Although  positive CC values are not natural in string theory, a mechanism for their appearance was provided in \cite{kklt}. Hence it is plausible to assume that the CC spectrum is sufficiently dense around zero such that the observed value is sufficiently close to some CC spectrum value.

The second part of the CC problem has been only addressed in string theory. This is the multiverse proposal, see \cite{mult}, and the assumption is that there are many universes, each having a fixed CC value from the CC spectrum. We live in the universe with the CC value $\L_c l_P^2 \approx 10^{-122}$, because this is the value that allows formation of galaxies, planets and life, see \cite{wein} for the anthropic determination of the CC value. 

Note that there are many proposals for P2 which are not derived from a QG theory, but instead it is assumed that a certain effective action exists such that its equations of motion give the required CC value, see for example \cite{nqgc}. 

\section{Conclusions}

The PLQG theory is a theory of quantum gravity which has finitely many degrees of freedom and no infinities. The underlying spacetime structure is a PL manifold $T(M)$ and the smooth spacetime $M$ is recovered as an approximation valid when the number of 4-simplices is large and at a length scale much larger than the typical edge length. The smooth spacetime approximation is analogous to the  smooth vector field approximation for the molecular velocities in a fluid.

The PLQG theory is defined by the Regge path integral with a non-trivial measure. The measure is chosen such that it gives a finite path integral, and also it has to admit a semi-classical solution of the effective action equation. These criteria select the exponentially vanishing measures for large edge lengths, and a simple and natural choice for the measure is (\ref{ccm}). This measure simplifies the analysis of the effective cosmological constant and one can obtain the formula (\ref{efcc}) for the exact effective CC, i.e. to all orders in $\hbar$. The two free parameters in (\ref{efcc}) can be consistently chosen such that the observed CC value is obtained. This is an important requirement for any QG theory and PLQG is the only existing QG theory where this property has been demonstrated explicitly.

Another nice property of the PLQG theory is that the effective action $\G$ can be approximated by a QFT effective action $\G^*$ when the number of 4-simplices in $T(M)$ is large. $\G^*$ can be calculated by using the perturbative QFT for GR with matter and with a momentum cutoff $K$, when $L_\e \ge L_K \gg l_P$. Hence the minimal edge length $L_K$ in the triangulation determines the momentum cutoff $K$ and
\be \G (L_1,\cdots, L_E ,\phi_1 ,\cdots,\phi_V ) \approx \G^*(g(x),\phi(x), K) \,, \label{qfta}\ee
for $E\gg 1$ and $V \gg 1$.

The QFT approximation (\ref{qfta}) will be still valid for $L_K \le l_P$, but in this case $\G^*$ cannot be calculated by the perturbative QFT methods. Instead, one has to use a non-perturbative method to solve the EA equation. The existence of the QFT approximation (\ref{qfta}) implies that one can obtain the running of the elementary particle masses and the coupling constants with $K$, see for example the equation (\ref{krun}). 

Note that the effective action only makes sense for the spacetimes which are given by the direct product of a 3-manifold with an interval. In order to study the quantum cosmology questions, one needs to consider 4-manifolds of general topology, which is different from $\S\times I$ topology. When $M \neq \S\times I$,  the concept of the effective action cannot be used. However, the Hartle-Hawking (HH) wavefunction \cite{HH} can be defined for any $T(M)$ by using the PLQG path integral (\ref{crss}). By choosing a triangulation for a manifold
$$ M \cup \left(\S \times I \right)\,,\quad \partial M = \S \,,$$
one can describe a Big-Bang quantum cosmology with an initial HH state, which evolves by the evolution operator defined by the PLQG path integral for the $T(\S\times I)$ part of the spacetime. It is then plausible to assume that the effective dynamics which corresponds to the time evolution of the HH state will be given by the PLQG effective action, defined by the equation (\ref{gmea}).

\end{document}